\documentclass[11pt]{article}
\usepackage{times}
\usepackage{geometry}
\usepackage[utf8]{inputenc}
\usepackage{enumitem,amssymb}
\usepackage{ragged2e}
\usepackage{graphicx}
\usepackage{comment}
\usepackage{multicol}
\usepackage{lineno}
\usepackage{bm}
\usepackage[usenames]{xcolor} %used for font color              
\definecolor{xlinkcolor}{cmyk}{1,1,0,0}
\usepackage{url}
\usepackage[
 colorlinks=true,    % false: boxed links; true: colored links 
 linkcolor=xlinkcolor,     % color of internal links            
 citecolor=xlinkcolor,     % color of links to bibliography
 filecolor=xlinkcolor,  % color of file links 
 urlcolor=xlinkcolor,      % color of external link
 final=true
]{hyperref}
\usepackage{enumitem}
\usepackage{xspace}
\usepackage[backend=biber,sorting=none, style=numeric-comp]{biblatex}
\newcommand{\topic}[1]{\noindent\textbf{#1}}

\setenumerate{itemsep=0mm}

\begin{document} 
%\linenumbers
\begin{raggedright} 
% part of template, but does not look good
\huge
Snowmass2021 - Letter of Interest \hfill \\[+1em]
\textit{Self-consistent approach for measuring the energy spectra and composition of cosmic rays and determining the properties of hadronic interactions at high energy} \hfill \\[+1em]
\end{raggedright}

\normalsize

\noindent {\large \bf Thematic Areas:} 

\noindent $\square$ (CF1) Dark Matter: Particle Like \\
\noindent $\square$ (CF2) Dark Matter: Wavelike  \\ 
\noindent $\square$ (CF3) Dark Matter: Cosmic Probes  \\
\noindent $\square$ (CF4) Dark Energy and Cosmic Acceleration: The Modern Universe \\
\noindent $\square$ (CF5) Dark Energy and Cosmic Acceleration: Cosmic Dawn and Before \\
\noindent $\square$ (CF6) Dark Energy and Cosmic Acceleration: Complementarity of Probes and New Facilities \\
\noindent $\blacksquare$ (CF7) Cosmic Probes of Fundamental Physics \\
\noindent $\blacksquare$ (Other) CompF1, EF06, EF07  \\

\noindent {\large \bf Contact Information:} \\
\noindent Jose Bellido (The University of Adelaide) [{\tt jose.bellidocaceres@adelaide.edu.au}] \\

\noindent {\large \bf Authors:} (institutions are provided at the end)

\noindent Andrea Addazi, Andy Buckley, Jose Bellido, CAO, Zhen, Ruben Concei\c{c}{\~a}o, Lorenzo Cazon, Armando di Matteo, Bruce Dawson, Kasumasa Kawata, Paolo Lipari, Analiza Mariazzi, Marco Muzio, Shoichi Ogio, Sergey Ostapchenko, M\'{a}rio Pimenta,  Tanguy Pierog, Andres Romero-Wolf, Felix Riehn, David Schmidt, Eva Santos, Frank Schroeder, Karen Caballero-Mora, Pat Scott, Takashi Sako,   Carlos Todero Peixoto, Ralf Ulrich, Darko Veberic, Martin White \\

\noindent {\large \bf Abstract:}

Air showers, produced by the interaction of energetic cosmic rays with the atmosphere, are an excellent alternative to study particle physics at energies beyond any human-made particle accelerator.  For that, it is necessary to identify first the mass composition of the primary cosmic ray (and its energy). None of the existing high energy interaction models have been able to reproduce coherently all air shower observables over the entire energy and zenith angle phase space. This is despite having tried all possible combinations for the cosmic ray mass composition. This proposal outlines a self-consistent strategy to study high energy particle interactions and identify the energy spectra and mass composition of cosmic rays. This strategy involves the participation of different particle accelerators and astrophysics experiments. This is important to cover the entire cosmic ray energy range and a larger phase-space of shower observables to probe the high energy interaction models.

\clearpage

% LOI text

Current high energy hadronic interaction models are not able to predict coherently all properties observed in energetic air showers~\cite{augerMPD2014,has_muons2015,auger_topdown2016,augerDelta2017,sigmaRmu_ICRC2019,auger_muons_umd2020}. Are the extrapolations of particle properties to higher energies incorrect? Are there any new particle physics phenomena at higher energies?  What really happens when energetic cosmic rays (protons or nuclei, $E>10^{18}$ eV) collide with nitrogen or oxygen nuclei at the top of the atmosphere? It is very important to highlight that collider physics might never access these energies.

The goal of this proposal is to use a range of  accelerator data in conjunction with observations from astophysical experiments  in order to determine the properties of particle interactions at energies up to $10^{19}$ eV. The astrophisical experiments considered are: Pierre Auger Observatory (Auger, Argentina)~\cite{longXmax2014}, Telescope Array (TA, USA)~\cite{Abbasi:2018wlq}, LHAASO (China)~\cite{zhen2010lhaaso}, HAWC~\cite{Abeysekara_2018} (Mexico), Yakutsk~\cite{yakutsk_composition} (Russia), IceCube/IceTop (South Pole)~\cite{ABBASI201315}, Tibet ASgamma (China)~\cite{HUANG201518} and ALAPCA~\cite{Calle_2020}, SWGO~\cite{Abreu:2019ahw} and TAMBO~\cite{Romero-Wolf:2020pzh} (South America). This project will benefit from undergoing detector upgrades (AugerPrime~\cite{prime_pdr2016, AugerPrime_ICRC2019, AugerRadio_ICRC2019} and TA$\times$4~\cite{tax4_uhecr2018}) and from low energy detector enhancements (HEAT and AMIGA~\cite{Klages:2012coa} and TALE~\cite{Ogio:2019Oc}) in Auger and TA. Archived data could be considered as well~\cite{Bellido:2018toz}. The expected outcomes include an enhanced capability to:
\begin{itemize}
  \item probe properties of particle interactions at energies well beyond the reach of particle colliders and
  \item determine energy spectra and mass composition of cosmic rays in order to evaluate scenarios for the origin of cosmic rays.
\end{itemize}
  This proposal envisages the development of new knowledge at the forefront of two important fields, Particle Physics and Astrophysics.
  The strength of this proposal is that the information collected by accelerator experiments and  astrophysical experiments will be used in a single analysis. By combining  the data from these detectors, hadronic interaction models will be constrained over an extensive energy range, from $10^{11}$ to $10^{19}$ eV. The  population of different types of particles in the air shower (i.e. muons and electromagnetic particles) will be measured at different energy ranges and at different atmospheric depths. The ratio between the muon and electromagnetic populations is closely tied to the cosmic ray composition and the particle interaction properties. For example, LHAASO ground detectors will sample air showers earlier in the atmosphere, since it is located at an altitude of 4400 m.a.s.l., AugerPrime and TA are located at 1400 m and IceCube/IceTop are located at 2835 m.a.s.l.. Other ground array experiments currently under design, such as ALPACA, SWGO and TAMBO, have the potential to contribute significantly to this project. The energy scale systematics from the experiments needs to be considered for the analysis~\cite{Cazon:2020zhx}. The more experiments operating within the same energy range, the better to reduce the effects of systematic uncertainties.   

  {\bf The GAMBIT collaboration} has previously developed a global and modular beyond-the-standard-model inference tool~\cite{Athron:2017ard, Kvellestad:2019vxm}, an open-source, modular package for performing global statistical fits of new particle physics theories with a broad range of collider and astrophysics data. The tool includes interfaces to state-of-the-art sampling algorithms adapted to both the Bayesian and frequentist statistical frameworks. Currently it includes a wide range of data from the Large Hadron Collider, dark matter direct and indirect searches, neutrino experiments, flavor physics, axion experiments and cosmological observations. In this project, we will extend it to include data from the experiments listed above, which can then be used in combined global fits with existing observables.

\begin{figure}[h]
\includegraphics[width=\textwidth]{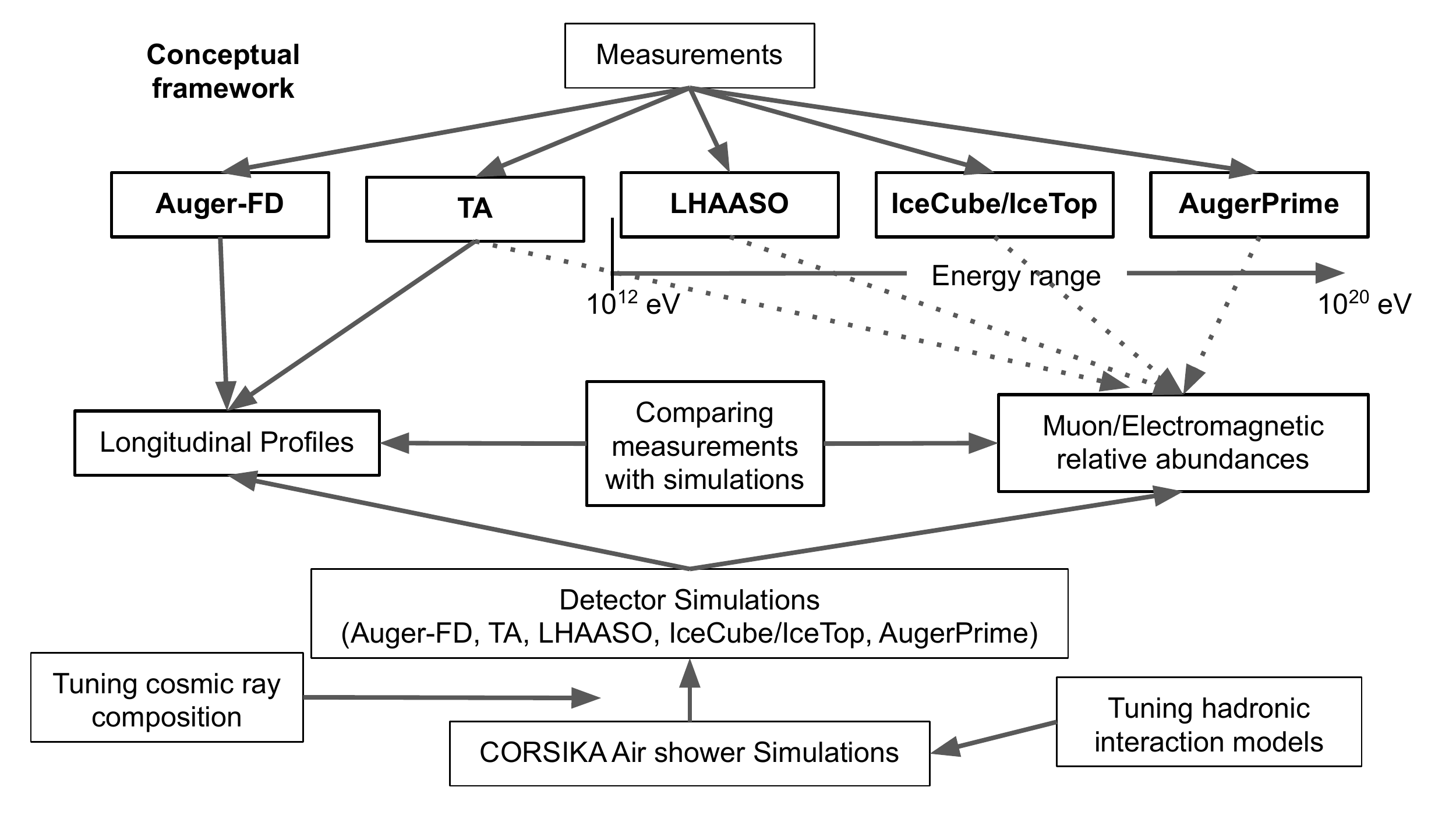}
\caption{Diagram of the conceptual framework. Data from particle accelerators and from other astrophysical experiments would or could be included in the project.}
\label{framework}
\end{figure}

\topic{Methodology}:

A diagram of the methodology is shown in Figure~\ref{framework}, and below is a step by step description.
\begin{enumerate}[label=\alph*)]

\item Measure the air shower lateral distribution (at the corresponding detector level) for the muonic and electromagnetic components. These measurements will be done for different cosmic ray zenithal angles.  Different ground arrays are optimized for different air shower energies and together they can cover energies ranging from $10^{11}$ eV to $10^{19}$ eV.
\item Measure the shower longitudinal profile using Cherenkov or fluorescence telescopes.

\item Using the package CORSIKA~\cite{Heck:1998vt}, perform simulations of air showers using each of the latest versions of popular high energy hadronic interaction models (e.g. QGSJET~\cite{refId0}, EPOS~\cite{Pierog:2013ria}, SIBYLL~\cite{Engel:2019dsg}). The simulated showers will need to cover the entire phase space of energies and zenithal angles as encountered in real air showers. A four-component mass composition of cosmic rays (p, He, N and Fe) could be simulated. The computational time for this task could be challenging despite using a supercomputer. However, we expect to share the simulation load between the different collaborations.   
  
\item Use the dedicated detector simulations for each experiment to simulate the detection of the simulated air showers (from step c). This step will generate the simulated data.

\item Repeat steps a) and b), but using the simulated data from step d).

\item Compare the observations from steps a) and b), with the corresponding expectations from step e) and characterize the differences. This information will provide valuable insights for improving high energy hadronic interaction models.
  
\item Explore modifications in the models of high-energy hadronic interactions in such a way that the simulation of an energy dependent composition mix of p, He, N and Fe,  match coherently with the observations in all experiments. The composition mix will change as a function of energy. The energy scale from each experiment will need to be normalized, so that all experiments will measure the same energy spectra and compositions (some considerations should be taken to study possible Northern/Southern sky differences).  

\end{enumerate}

It is important to point out that there exist direct measurements (from satellite and balloon borne detectors) of  proton and Helium cosmic ray fluxes up to $10^{12}$ eV~\cite{Aguilar:2015ooa} and to up to $10^{13}$ eV for nuclei. These measurements could be used to validate the above methodology at the lowest energy.

\clearpage

\printbibliography[title={References}]
%\noindent {\large \bf References:} (hyperlinks welcome)
%\printbibitembibliography
%\input{LoI_bib2bib.bbl}

%\vspace{4in}

%\clearpage

\noindent {\large \bf Authors:} (names and institutions)

\noindent Andrea Addazi$^1$, Andy Buckley$^2$, Jose Bellido$^3$, CAO, Zhen$^4$, Ruben Concei\c{c}{\~a}o$^5$, Lorenzo Cazon$^5$, Armando di Matteo$^6$, Bruce Dawson$^3$, Kazumasa Kawata$^7$, Paolo Lipari$^8$, Analiza Mariazzi$^9$,  Marco Muzio$^{10}$, Shoichi Ogio$^{11}$, Sergey Ostapchenko$^{12,13}$, M\'{a}rio Pimenta$^5$,  Tanguy Pierog$^{14}$, Andres Romero-Wolf$^{15}$,  Felix Riehn$^5$, David Schmidt$^{14}$, Eva Santos$^{16}$, Frank Schroeder$^{17}$, Karen Caballero-Mora$^{18}$, Pat Scott$^{19}$, Takashi Sako$^{7}$, Carlos Todero Peixoto$^{20}$, Ralf Ulrich$^{14}$, Darko Veberic$^{14}$, Martin White$^3$ \\

\noindent $^1$Sichuan University, China\\
$^2$ University of Glasgow, UK\\ 
$^3$The University of Adelaide, Australia\\
$^4$Institute of High Energy Physics, China\\
$^5$Laboratory of Instrumentation and Experimental Particle Physics, Portugal\\
$^6$Istituto Nazionale di Fisica Nucleare, Italy\\
$^{7}$University of Tokyo, Japan \\
$^8$Sapienza University of Rome, Italy\\
$^9$Universidad Nacional de La Plata, Argetina\\
$^{10}$New York University, USA\\
$^{11}$Osaka City University, Japan\\
$^{12}$Frankfurt Institute for Advanced Studies, Germany\\
$^{13}$Moscow State University, Russia\\
$^{14}$Karlsruhe Institute of Technology, Germany\\
$^{15}$Jet Propulsion Laboratory, California Institute of Technology, USA\\
$^{16}$Institute of Physics of the Czech Academy of Sciences, Czech Republic\\
$^{17}$University of Delaware, USA\\
$^{18}$Universidad Aut\'onoma de Chiapas, M\'exico\\
$^{19}$The University of Queensland, Australia \\
$^{20}$ University of Sao Paulo, Brazil \\

\end{document}